\documentclass[prb, reprint]{revtex4-1}
\usepackage{hyperref}
\usepackage{geometry}                
\usepackage{amsmath}
\usepackage{bm}
\usepackage{graphicx}
\usepackage{natbib}
\usepackage{subfigure}
\usepackage{amssymb}
\usepackage{epstopdf}
\usepackage{color}
\begin{document}

\title{ Excitonic Instabilities and Insulating States in Bilayer Graphene}
\author{Kok Wee \surname{Song}}
\author{Yung-Ching \surname{Liang}}
\author{Stephan \surname{Haas}}

\affiliation{Department of Physics and Astronomy, University of Southern California, California 90089 USA}

\begin{abstract}
The competing ground states of bilayer graphene are studied by applying renormalization group techniques to a bilayer honeycomb lattice with nearest neighbor hopping. In the absence of interactions, the Fermi surface of this model at half-filling consists of two nodal points with momenta $\mathbf{K}$, $\mathbf{K}'$, where the conduction band and valence band touch each other, yielding a semi-metal. Since near these two points the energy dispersion is quadratic with perfect particle-hole symmetry,  excitonic instabilities are inevitable if inter-band interactions are present. Using a perturbative renormalization group analysis up to the one-loop level, we find different competing ordered ground states, including ferromagnetism, superconductivity, spin and charge density wave states with ordering vector $\mathbf{Q}=\mathbf{K}-\mathbf{K}'$, and excitonic insulator states. In addition, two states with valley symmetry breaking are found in the excitonic insulating and ferromagnetic phases. This analysis strongly suggests that the ground state of bilayer graphene should be gapped, and with the exception of superconductivity, all  other possible ground states are insulating.
\end{abstract}

\maketitle


\section{Introdution}

\begin{figure}
\begin{center}
\includegraphics[scale=0.2]{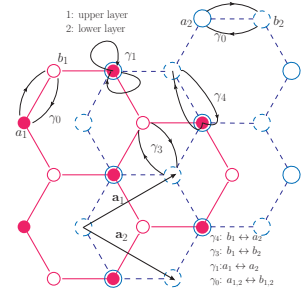}
\end{center}
\caption{\label{BLG} Bilayer graphene with AB-stacking: $a_1$, $b_1$ are the two sublattice sites in the upper layer, $a_2$, $b_2$ are the two sublattice sites in the lower layer. $\gamma_0$ is the tight-binding hopping constant between $a_1$; $b_1$, $\gamma_1$ is the hopping between $a_1$ and $a_2$; $\gamma_3$ is the hopping between $b_1$ and $b_2$. $\mathbf{a}_1=\frac{a}{2}(3,\sqrt3)$ and $\mathbf{a}_2=\frac{a}{2}(3,-\sqrt3)$ are the primitive lattice vectors.}
\end{figure}

Graphene is a quasi-2D carbon material with a honeycomb lattice structure. Its band structure is captured by a tight binding model, as illustrated in Fig. \ref{BLG}, with two interpenetrating triangular sublattices $a$ and $b$ 
\begin{equation*}\label{graphene}
H_{A}= -\gamma_0\sum_{\langle i,j\rangle}a^{\dagger}_i b_j+h.c.,
\end{equation*}
where $\langle i,j\rangle$ denotes a sum over all nearest neighbor pairs. At the charge neutrality point, this model yields a semi-metal for which the Fermi surface (FS) contains only two nodal points. Since the energy dispersion is linear in the vicinity of these  Dirac points, the corresponding low-energy effective Hamiltonian is given by a 2D Dirac model. This unique electronic structure leads to many interesting phenomena.\cite{RevModPhys.81.109} 

Although interactions between electrons are present in graphene, the one-particle picture works surprisingly well. In contrast to ordinary metals,  the ground state of the electrons in graphene does not behave like a Landau Fermi liquid, but rather belongs to the universality class of Dirac liquids.\cite{Sheehy:2007ys} One of the differences between these ground states is that short-range interactions between electrons are irrelevant in Dirac liquids.\cite{2011RSPTA.369.2625V} This may explain why the one-particle picture is applicable, regardless of the perfect particle-hole nesting properties of the lattice. However, recent experiments have shown evidence that the Dirac cone is renormalized,\cite{Elias:2011ly} suggesting that electron interactions are important on some level. Recently, the interactions between electrons in graphene have been modeled by a long-range Coulomb interaction or by using an effective (2+1)D QED model.\cite{2011arXiv1112.2054C,2010arXiv1012.3484K,2011RSPTA.369.2625V}

For bilayer graphene (BLG), tight-binding calculations also show that the non-interacting ground state  is a semi-metal. But in this case, the dispersion near the FS points is quadratic rather than linear.\cite{PhysRevLett.96.086805} Because of this, all short-range interactions now become relevant perturbations, and recent theories have predicted various possible spontaneous symmetry breaking ground states.\cite{PhysRevB.77.041407,PhysRevLett.104.156803,Vafek:2010kx,Zhang:2010vn,PhysRevB.82.201408,2011arXiv1111.2076T,2012arXiv1206.0288C,PhysRevB.82.115124}Furthermore, recent experiments\cite{Mayorov12082011,VelascoJ.:2012fk,PhysRevLett.105.256806,Weitz05112010,PhysRevLett.108.076602} have shown some evidence for FS reconstruction in BLG. These findings contradict the simple one-particle picture for BLG, based on a tight-binding model, and rather suggest that interactions between electrons play an important role in breaking down the FS.

In this paper, the instabilities in BLG will be addressed by using a perturbative renormalization group approach. We consider the bilayer honeycomb structure with nearest neighbor hopping as the  low-energy effective model  for BLG. Particle-hole symmetry is assumed, and RG arguments are used to identify the dominant channels and eliminate the irrelevant channels due to the interactions in the model. Using this setup, an array of possible ordered phases is found, which are competing with  each other. In the following sections, the details of the model and the results and implications of our calculations will be discussed.

\section{Bilayer Graphene and the model Hamiltonian}

\begin{figure}
\begin{center}
\subfigure[First Brillouin Zone]{\includegraphics[scale=0.3]{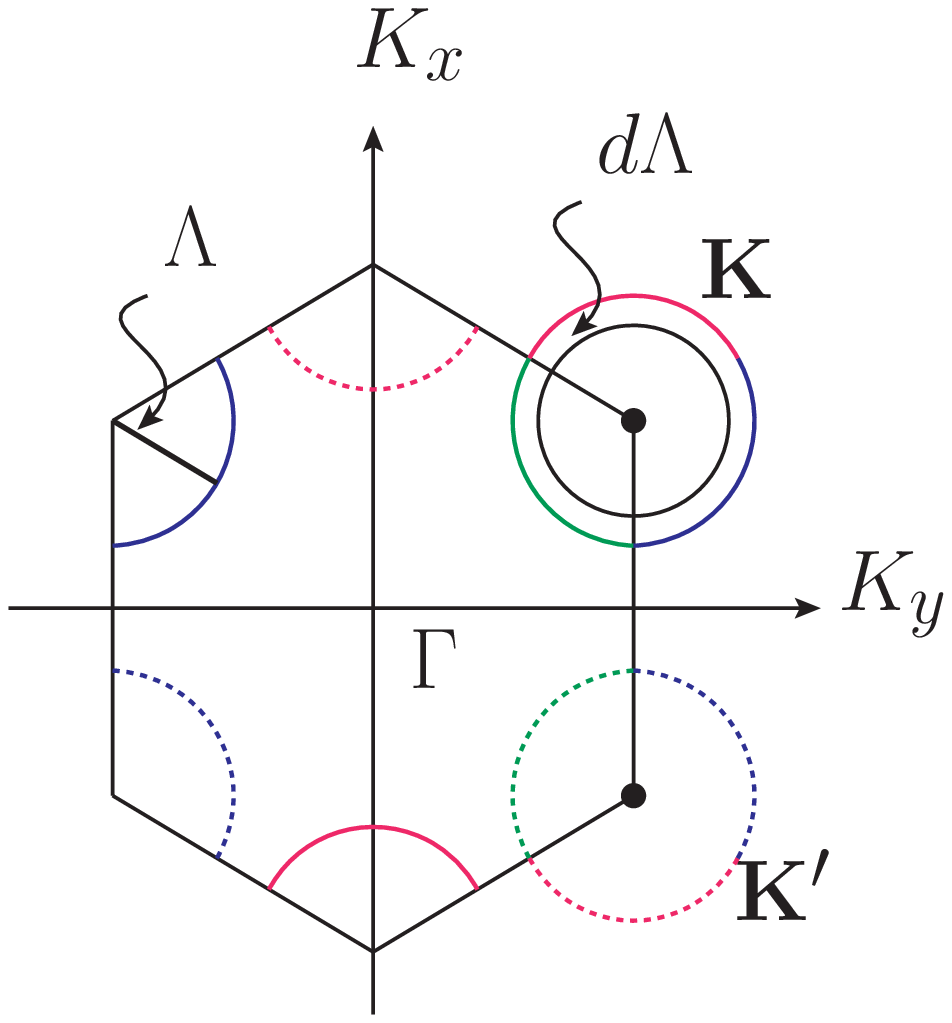}\label{1BZ}} 
\subfigure[Energy dispersion around the Fermi points]{\includegraphics[scale=0.3]{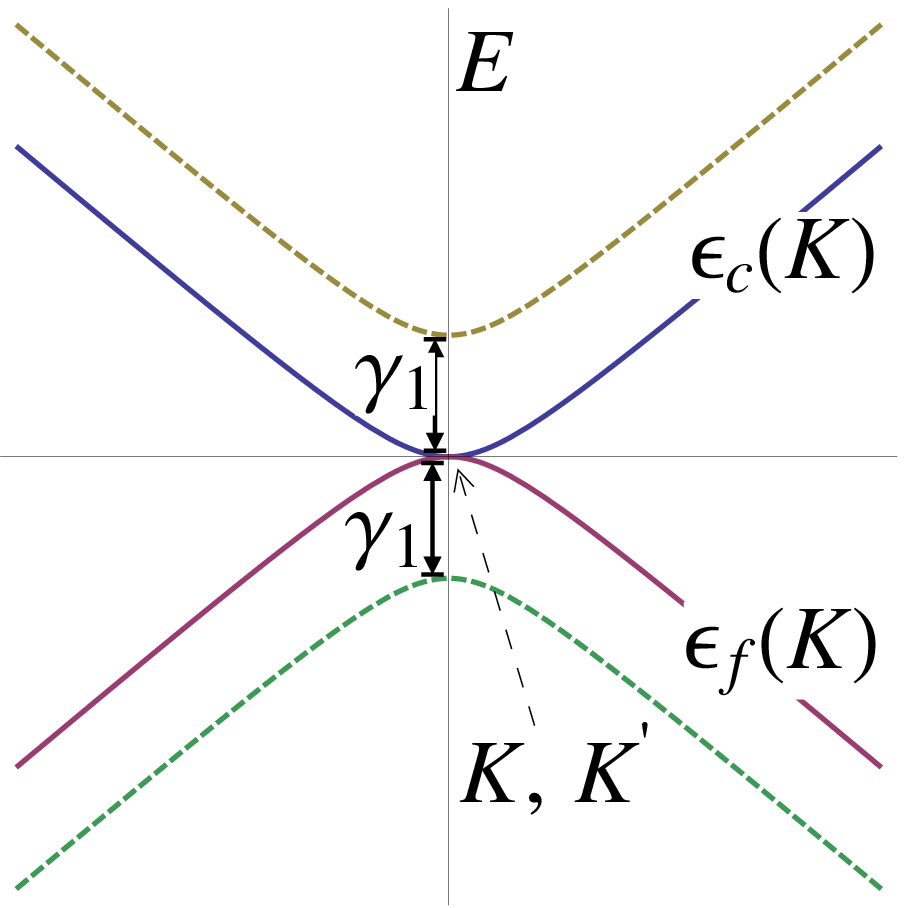}\label{Band}}
\end{center}
\caption{(a) $\mathbf{K}=\frac{2\pi}{3a}(1,\frac{1}{\sqrt3})$ and $\mathbf{K^\prime}=\frac{2\pi}{3a}(1,-\frac{1}{\sqrt3})$ are the two points,
constituting the Fermi surface of the non-interacting system. $\Lambda\leq\gamma_1$ is the energy cutoff of the theory, $d\Lambda$ is a thin shell contain high energy modes to be integrated out. (b) $\epsilon_{c}(K)$ and $\epsilon_{f}(K)$ are the dispersion energy of the conduction and the valence band respectively. The other two bands are gapped by $\gamma_1$.}
\end{figure}

The crystal structure of BLG  is given by  a Bernal AB stacking of two sheets of graphene, shown in Fig. \ref{BLG}). In the absence of interactions, its band structure is effectively described by a tight-binding model.\cite{RevModPhys.81.109} In momentum space, the one-particle Hamiltonian with $\gamma_4\simeq0$ is given by
\begin{equation*}
H_{AB}=\sum_{K,\sigma}\Psi^\dagger_{K\sigma} \mathcal{H}_K \Psi_{K\sigma},
\end{equation*}
where $\mathcal{H}_K $ is 
\begin{equation}\label{HABK} 
\begin{pmatrix}  
	0		& \gamma_0 f(K) & 		0	 & \gamma_3 f^\ast(K) \\
 \gamma_0 f^\ast(K) & 	0 		&\gamma_1 & 		0		\\
 	0 		& \gamma_1 	& 		0 	& \gamma_0 f(K) \\
\gamma_3 f(K) & 	0 		&\gamma_0f^\ast(K) & 		0		
\end{pmatrix},
\end{equation}
$\Psi^\dagger_{K\sigma}=\bigl(b_{1K\sigma}^\dagger, a_{1K\sigma}^\dagger, a_{2K\sigma}^\dagger, b_{2K\sigma}^\dagger\bigr)$, $f(K)=\sum_{i=1}^3 e^{iK\cdot\delta_i}$ are the orbital field operators, and
$\delta_1=\frac{a}{2}(1,\sqrt3)$, $\delta_2=\frac{a}{2}(1,-\sqrt3)$, $\delta_3=a (-1,0)$ are nearest-neighbor in-plane displacement vectors ($a$ is the lattice constant). Fig. \ref{1BZ} shows the 1st Brillouin zone in momentum space with reciprocal vectors $b_1=\frac{2\pi}{3a}(1,\sqrt3)$ and $b_2=\frac{2\pi}{3a}(1,-\sqrt3)$.

Since only low energy excitations are of  interest  here, we expand $f(K)$ near $\mathbf{K}$ and $\mathbf{K'}$ (up to a phase factor $e^{i\pi/6}$),
\begin{equation*}
f(K)\simeq-\frac{3a}{2}\kappa\text{ at }\mathbf{K}, \quad f(K)\simeq-\frac{3a}{2}\kappa^\ast\text{ at }\mathbf{K'},
\end{equation*}
where $\kappa=k_x+ik_y$, $k=(k_x,k_y)$ is a small momentum deviation from $\mathbf{K}$, $\mathbf{K^\prime}$, and $|k|\leq\Lambda\ll|\mathbf{K}|,|\mathbf{K'}|$.

In the following discussion, the trigonal warping term $\gamma_3$ will be neglected. (The justification for this will be discussed in the Sec. \ref{discussion}). As shown in Fig. \ref{Band}, the resulting tight-binding band structure consists of 4 bands. Two of these bands are gapped by $\gamma_1$ from the FS, whereas the other two bands touch each other at the $\mathbf{K}$ and $\mathbf{K'}$ Fermi points. This is similar to single-layer graphene, but for the bi-layer case the energy dispersion is quadratic at the Fermi surface,
\begin{equation}\label{dispersion}
\epsilon_{c,f}(K)\simeq\pm\frac{v_F^2}{\gamma_1}k^2 \quad \text{at }\mathbf{K}, \mathbf{K'}.
\end{equation}

In the following analysis of instabilities, the gapped bands will be ignored, because they are not important in the low energy limit. Before writing down the model Hamiltonian, let us introduce the creation (annihilation) operators for electrons in bands $\epsilon_{c}(K)$ and $\epsilon_{f}(K)$ to be $c_{K\sigma}$ ($c^\dagger_{K\sigma}$) and $f_{K\sigma}$ ($f^\dagger_{K\sigma}$) respectively. $c_{K\sigma}$ and $f_{K\sigma}$ are linear combinations of the local orbital field operators ($b_{1K\sigma}, a_{1K\sigma}, a_{2K\sigma}, b_{2K\sigma}$):
\begin{equation}\label{Eigenstate}
c_{K\sigma }= \mathbf{C}^c_{K}\cdot \Psi_{K\sigma},\quad f_{K\sigma }=\mathbf{C}^f_{K}\cdot \Psi_{K\sigma},
\end{equation}
where  $\mathbf{C}^I_{K}=(C^I_{1K}, C^I_{2K},C^I_{3K},C^I_{4K})$.  These coefficients near the Fermi surface can be found in Ref. \onlinecite{PhysRevB.73.214418}. Note that if the model is written using the local orbital basis, in momentum space these coefficient $\mathbf{C}^I_{K}$ account for the form factors in the interaction terms of the Hamiltonian. Because of the small $k$ dependence in the form factors, this can complicate the RG analysis. In order to avoid this problem, it is natural using the Bloch wave basis to build an effective Hamiltonian of BLG. Then, the non-interacting part of the model Hamiltonian can be represented as
\begin{equation}\label{H0}
H_0=\sum_K \epsilon_c(K)c^{\dagger}_{\sigma K}c_{\sigma K}+ \epsilon_f(K)f^{\dagger}_{\sigma K}f_{\sigma K},
\end{equation}
where summing over all $\sigma$  is implicitly assumed.

Turning to the interaction part of the Hamiltonian, we follow the approach outlined in Ref.\onlinecite{PhysRevB.84.205445} (for more details see Appendix \ref{Ainteractions}), and require particle-hole symmetry of exchanging the valence and conduction bands. Then the  electron-electron interaction term can be written as
\begin{equation}\label{Hint}\begin{split}
&H_{int} =\frac{1}{2}\sum_{\substack{K_1,K_2,\\ K_3,K_4}} \\
&\{\;\;\, \mathcal{U}_0(K_3K_4K_2K_1)c^{\dagger}_{K_3\sigma }c^{\dagger}_{K_4\sigma'}c_{K_2\sigma'}c_{K_1\sigma }\\
&+\mathcal{U}_1(K_3K_4K_2K_1)c^{\dagger}_{K_3\sigma }c^{\dagger}_{K_4\sigma'}f_{K_2\sigma'}f_{K_1\sigma}\\
&+\mathcal{U}_2(K_3K_4K_2K_1)c^{\dagger}_{K_3\sigma}f^{\dagger}_{K_4\sigma'}f_{K_2\sigma' }c_{K_1\sigma}\\
&+ \mathcal{U}_3(K_3K_4K_2K_1)f^{\dagger}_{K_3\sigma }c^{\dagger}_{K_4\sigma'}f_{K_2\sigma'}c_{K_1\sigma}
\}\\
&+\{\text{exchange } (c\leftrightarrow f)\},
 \end{split}\end{equation}
where momentum conservation is implicitly contained in $\mathcal{U}$ (see Appendix \ref{Ainteractions}). Here, the coupling constant $\mathcal{U}_0$ denotes the intra-band interaction, whereas $\mathcal{U}_1$, $\mathcal{U}_2$ and $\mathcal{U}_3$ are inter-band interactions. 

So far, no explicit advantages are obvious by using the Bloch wave basis. In addition, the momentum dependence in the coupling constants complicates the study too. However, this complication will be removed due to the trivial topology of the FS. (See below)

\section{Renormalization Group Analysis of The BLG Model}

Here, we apply the  pertubative renormalization group (RG) method to explore the low-energy physics of the BLG model in the presence of interactions, following the standard procedure outlined in  Ref. \onlinecite{Shankar:1994uq}. 

From a tree level analysis (see Appendix \ref{RGTree}), we find that only a finite set of coupling constants are marginal. In the low energy limit, only the interacting channels which depend on $\mathbf{K}$,$\mathbf{K^\prime}$ are not renormalized to zero. The corresponding bare coupling constants are listed and classified into Table \ref{coupling}.

\begin{table}[h]
  \centering
  \begin{tabular}{@{} |c|cccc| @{}}
  \hline
    & $\mathcal{U}_0$ &   $\mathcal{U}_1$   &   $\mathcal{U}_2$   &   $\mathcal{U}_3$  \\ 
    \hline
    $\mathcal{U}(\mathbf{K}$,$\mathbf{K}$,$\mathbf{K}$,$\mathbf{K})$& $h_0$ & $g_0$ & $u_0$ & $v_0$ \\ 
   $\mathcal{U}(\mathbf{K}$,$\mathbf{K'}$,$\mathbf{K'}$,$\mathbf{K})$& $h_1$ & $g_1$ & $u_1$  & $v_1$\\ 
   $\mathcal{U}(\mathbf{K'}$,$\mathbf{K}$,$\mathbf{K'}$,$\mathbf{K})$& $h_2$ & $g_2$ & $u_2$  & $v_2$\\ 
  \hline
  \end{tabular}
  \caption{Bare coupling parameters of the marginally relevant processes.}
  \label{coupling}
\end{table}

Here, the subscripts $0$, $1$, $2$ of the coupling constants indicate the various scattering processes between valleys. The difference between processes with $0$, $1$ versus  $2$ is that after scattering processes with $0$, $1$ do not exchange valley indices between two particles, but processes with $2$ do. Therefore, the scattering processes with subscript $2$ always involve large momentum transfers.

Since these coupling constants are  marginal, performing one-loop corrections to the RG flow equations is necessary. Since the interaction is quartic, i.e. involving only two-body scattering, there are only three distinct channels to transfer momentum. Following the terminology of Ref. \onlinecite{Shankar:1994uq}, these processes are named ZS, ZS', and BCS.

\begin{figure}[h]
\begin{center}
\scalebox{0.3}{\includegraphics{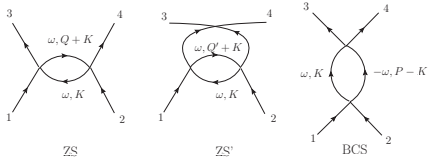}}
\end{center}
\caption{\label{Feynman} Feynman Diagrams: $1,2,3,4$ represent the low energy modes with momentum $K_{1,2,3,4}$, band index $I$, valley index $\alpha$, and spin $\sigma$. The momentum inside the loop, $K$, must lie within the shell $d\Lambda$, and $Q=K_3-K_1$, $Q^\prime=K_4-K_1$, $P=K_1+K_2$. Note that the interaction lines are suppressed.}
\end{figure}
 
The corresponding Feynman diagrams are schematically shown in Fig. \ref{Feynman}. All modes in the loop are high energy and need to be integrated out. After rescaling back to the original phase space volume, the coupling constants are modified, i.e. they are flowing in a 12 dimensional space of couplings. 

In order to have non-vanishing one-loop corrections, in the ZS and ZS' diagrams the two propagators in the loop must pair up with a different band. For BCS, both propagators must pair up within the same band. Those graphs that do not satisfy the above criteria contain double poles in the frequency $\omega$ contour integration. With this, many contributions of these diagram can be eliminated, thus greatly simplifying the calculation. 

In this work, we consider flow equations for the couplings up to the one-loop level. Cumulant expansion and  Wick's contraction are used in the calculation. This method is convenient to keep track of the prefactor for each different diagram.

The loop momentum integration (bubble diagram) can be evaluated,
\begin{equation}
\int^{2\pi}_0\int^{\Lambda}_{\Lambda-d\Lambda}\frac{d\theta kdk}{(2\pi)^2} \frac{1}{2|\epsilon_{I}(K)|}=\frac{dt}{4\pi\nu_0},
\end{equation}
where $\nu_0=v_F^2/\gamma_1$, and $dt=\frac{d\Lambda}{\Lambda}$ is the RG running parameter. Therefore, the RG flow rate equations under one-loop correction are given by,
\begin{widetext}
\begin{eqnarray}
&\frac{d}{dt}
\left[
\begin{array}{c}
  h_0   \\
  h_1   \\
  h_2   
\end{array}
\right]
=
\frac{1}{4\pi\nu_0}\left[
\begin{array}{c}
  -h_0^2-g_0^2 \\
  -h_2^2-h_1^2 -g_2^2-g_1^2\\
  -2h_1h_2 -2g_1g_2
\end{array}
\right]\label{RGflowh};
\frac{d}{dt}
\left[
\begin{array}{c}
  u_0   \\
  u_1   \\
  u_2   
\end{array}
\right]
=
\frac{1}{4\pi\nu_0}\left[
\begin{array}{c}
  u_0^2+u_2^2+g_0^2+g_2^2  \\
  u^2_1 +g_1^2  \\
  2u_0u_2  + 2g_0g_2
\end{array}
\right]\label{RGflowu}\nonumber \\
&\frac{d}{dt}
\left[
\begin{array}{c}
  g_0   \\
  g_1   \\
  g_2   
\end{array}
\right]
=
\frac{1}{4\pi\nu_0}\left[
\begin{array}{c}
  -2g_0h_0-2g_0v_0+2g_2v_1-4g_1v_1+2g_0u_0+2g_2u_2+2u_2g_1\\
   -2g_2h_2-2g_1h_1-2g_0v_1-4g_1v_0+2g_2v_0+2g_1u_0+2g_0u_2+2g_1u_1\\
   -2g_1h_2-2g_2h_1-4g_2v_2+2g_1v_2+2g_0u_2+2g_2u_0+2g_2u_1
\end{array}
\right]\label{RGflowg} \nonumber\\ 
&\frac{d}{dt}
\left[
\begin{array}{c}
  v_0   \\
  v_1   \\
  v_2   
\end{array}
\right]
=
\frac{1}{4\pi\nu_0}\left[
\begin{array}{c}
  2(u_0-v_0)v_0+2(u_2-v_1)v_1 +2(g_2-g_1)g_1 \\
  2u_2v_0+2(u_0-2v_0)v_1+2(g_2-g_1)g_0\\
  2(u_1-v_2)v_2+2(g_1-g_2)g_2
\end{array}
\right]\label{RGflow}
\end{eqnarray}
\end{widetext}

If $g_0=g_1=g_2=0$, these RG flow rate equation can be solved exactly, and decoupled into a simple result,
\begin{widetext}
\begin{eqnarray}
&\frac{d}{dt}
\left[
\begin{array}{c}
  h_0   \\
  h_1+h_2   \\
  h_1-h_2 
\end{array}
\right]
=
\frac{1}{4\pi\nu_0}\left[
\begin{array}{c}
  -h_0^2 \\
  -(h_1+h_2)^2 \\
   -(h_1-h_2)^2
\end{array}
\right]\label{RGflowh0} ; 
\frac{d}{dt}
\left[
\begin{array}{c}
  u_0+u_2   \\
  u_1   \\
  u_0-u_2   
\end{array}
\right]
=
\frac{1}{4\pi\nu_0}\left[
\begin{array}{c}
  (u_0+u_2)^2  \\
  u^2_1\\
  (u_0-u_2)^2  
\end{array}
\right]\label{RGflowu0} \nonumber\\
&\frac{d}{dt}
\left[
\begin{array}{c}
  (u_0-2 v_0)+(u_2-2 v_1)\\
  (u_0-2 v_0)-(u_2-2 v_1)\\
  u_1-2v_2   
\end{array}
\right]
=
\frac{1}{4\pi\nu_0}\left[
\begin{array}{c}
  ((u_0-2 v_0)+(u_2-2 v_1))^2 \\
  ((u_0-2 v_0)-(u_2-2 v_1))^2\\
  (u_1-2v_2)^2\\ 
\end{array}
\right]\label{RGflowv0}
\end{eqnarray}
\end{widetext}

Before finishing this section, we need to address the effects of quadratic perturbations. The two most relevant perturbations are the chemical potenal and trigonal warping, i.e. the $\gamma_3$ hopping term. These perturbations are in principle relevant under the tree level, i.e. scaling as $s^2$ and $s$ respectively. The chemical potential determines the density of the system, and the trigonal warping splits the original two Fermi points into four.

However, the divergences of the susceptibilities (see next section) emerge at some finite energy scale, and the RG flow must be stopped at this point. This energy scale determines the ordered state mean field transition temperature $T_c$. If $T_c$ is far above the trigonal warping reconstruction energy, this quadratic perturbation is not significant. This introduces an infrared cut-off to the validity of the analysis. \cite{Vafek:2010kx,Zhang:2010vn,Zhang:2010vn,PhysRevB.82.201408}. In addition, the divergences also imply that the original FS is unstable towards opening a gap. Although one should follow the procedure in Ref. \onlinecite{Shankar:1994uq} to fine-tune the chemical potential to keep the system density fixed, not carrying out this procedure does not affect the results significantly. 

Because of this, we argue that trigonal warping and the chemical potential do not play an essential role in the analysis at the one-loop level, as long as the energy scale of the instabilities is found to be far beyond the infrared limit.

\section{Susceptibilities and Possible Ground States}

To gain further understanding into the physics of BLG, we introduce test vertices into the original Hamiltonian. \cite{PhysRevB.78.134512,Nandkishore:2012ve} These test vertices  correspond  to the pairing susceptibilities, 
\begin{eqnarray}
&\Delta_j\sum_{\alpha k} c^\dagger_{\alpha s, k}\tau^\mu_{\alpha\alpha'}\otimes\sigma^\nu_{ss'}f_{\alpha' s',k},\label{pairph}\\
&\Delta_{SC}\sum_{\alpha k}\begin{bmatrix} c_{\alpha s, k}\tau^x_{\alpha\alpha'}\otimes\sigma^y_{ss'}c_{\alpha' s',-k}\\+f_{\alpha s, k}\tau^x_{\alpha\alpha'}\otimes\sigma^y_{ss'}f_{\alpha' s',-k}\end{bmatrix}\label{pairpp1}\\
&\Delta_{SC'}\sum_{\alpha k}\begin{bmatrix} c_{\alpha s, k}\tau^x_{\alpha\alpha'}\otimes\sigma^y_{ss'}c_{\alpha' s',-k}\\-f_{\alpha s, k}\tau^x_{\alpha\alpha'}\otimes\sigma^y_{ss'}f_{\alpha' s',-k}\end{bmatrix}\label{pairpp2}
\end{eqnarray}
where, $\mu,\nu=0,x,y,z$, $\tau^0=\sigma^0$ are $2\times2$ identity matrices, $\tau^{x,y,z}$ and $\sigma^{x,y,z}$ are $2\times2$ Pauli matrices. $\tau$ denotes the  valley degree of freedom with basis $(\mathbf{K}, \mathbf{K'})$, and $\sigma$ denotes the spin degree of freedom. $j$ indicates the different pairings listed in Table \ref{Gamma}.

Performing an RG analysis at the one-loop level with this additional new perturbed Hamiltonian, the vertices ($\Delta$s) are renormalized, and the new renormalized vertices are of the form
\begin{equation}
\Delta_{j}^{Ren}=\Delta_{j}(1+\frac{1}{4\pi\nu_0}\Gamma_{j}\ln s),
\end{equation}
where the $\Gamma_j$ are listed in Table \ref{Gamma}.
\begin{table*}[t] 
  \centering
  \begin{tabular}{@{} |c|l|c|c|@{}}
  \hline
  $j$ & Ordered State &$\tau^\mu\otimes\sigma^\nu$& $\Gamma_j$ \\     \hline
    $FM$&ferromagnetism &$\tau^0\otimes\sigma^z$& $u_0+ g_0+u_2+g_2$\\  
     \hline 
    $FM'$ &FM without valley symmetry &$\tau^z\otimes\sigma^z$&$u_0+ g_0-(u_2+ g_2)$\\ 
    \hline
   $SDW$&spin density wave &$\tau^x\otimes\sigma^z$& $u_1+ g_1$   \\
       \hline
    $EI$&excitonic insulator &$\tau^0\otimes\sigma^0$& $(u_0-2v_0)+(u_2-2v_1)+ (g_0-(g_2 -2g_1)) $\\  
    \hline 
    $EI'$ & EI without valley symmetry &$\tau^z\otimes\sigma^0$&$(u_0-2v_0)-(u_2-2v_1)+ (g_0+(g_2 -2g_1))$ \\ 
   \hline
   $CDW$&charge density wave &$\tau^x\otimes\sigma^0$& $u_1+ g_1-2(v_2+ g_2)$ \\
      \hline
     $SC$&superconductor &$\tau^x\otimes\sigma^y$& $-(h_1-h_2+(g_1-g_2))$ \\
      \hline
      $SC'$&superconductor &$\tau^x\otimes\sigma^y$& $-(h_1-h_2-(g_1-g_2))$ \\
      \hline
  \end{tabular}
  \caption{LPairing susceptibilities corresponding to the competing ground states in the presence of interactions.}
  \label{Gamma}
\end{table*}

\subsection{Case I: $g_0=g_1=g_2=0$}

If $g_0=g_1=g_2=0$, Eq. \eqref{RGflowh0} decouples the susceptibilities,  and one obtains
\begin{equation}
\Gamma^{g=0}_j(t)=\frac{\Gamma^{g=0}_j(0)}{1-\frac{1}{4\pi\nu_0}\Gamma^{g=0}_j(0)t}\text{ .}
\end{equation}

Whether and where the susceptibilities ($\Gamma_j\ln s$, where $t=\ln s$)  diverge is  determined by the bare coupling constants. Each divergence in $\Gamma^{g=0}_j(t)$ indicates that the system has a tendency toward the corresponding ordered state, labeled by 'j'. The first instability in a given channel represents the most dominant ordered state of the system at low energy.

For the case $g_0=g_1=g_2=0$, the situation is relatively simple. If only repulsive interactions are considered, $FM$ and $SDW$ are the dominant instabilities. In order to produce instabilities in the other channels, fine-tuning of the bare parameters is needed. The $\Gamma_j(0)$ must be positive such that other mean field solutions exist. More generally, since the parameter space spanned by the bare couplings is very large, constraining the search is desirable in order to make the exploration and analysis of the phase diagram meaningful. 

\begin{figure}[h]
  \begin{center}
\includegraphics[scale=1.05]{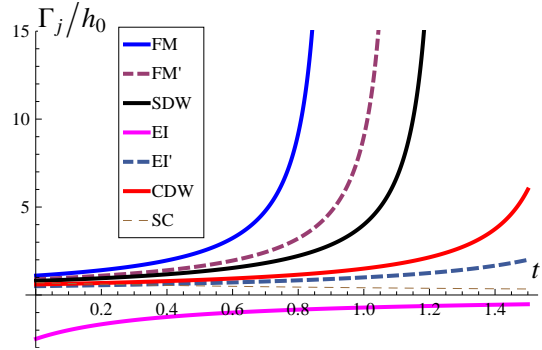}
  \end{center}
   \caption{Flow of the susceptibilities: Here, we set $h_0=u_0=v_0$, $h_1=u_1=v_1=0.8h_0$,  $h_2=u_2=v_2=0.1h_0$, and $g_0=g_1=g_2=0$ ($h_0>0$) . The $FM$ instabilities occupy a large region in parameter space, and fine-tuning is not necessary.} \label{Gammaflow1}
\end{figure}

The relative strength between the bare couplings can be estimated. In general, the scattering processes within the same valley $h_0$, $g_0$, $u_0$, $v_0$ and $h_1$, $g_1$, $u_1$, $v_1$ are expected to be larger than $h_2$, $g_2$, $u_2$, $v_2$, because  intra-valley scattering processes involve only small momentum transfer.\cite{doi:10.1146/annurev-conmatphys-020911-125055} Applying these constraints, in Fig. \ref{Gammaflow1} we show how these pairing susceptibilities compete with each other for a representative choice of bare coupling parameters. In this example, the dominant low-energy divergence occurs in the $FM$ channel, followed by $FM'$, $SDW$, $CDW$ and $SC$ at higher energy scales. 

The instabilities $FM$, $FM'$, and $SDW$ indicate broken spin symmetry, thus leading to magnetically ordered ground states. Since the pairing in \eqref{pairph} is a pairing of different bands $c_\sigma$, $f_\sigma$, it does not have an obvious connection with the spin density operator. However, it can be related to local magnetization in a more sophisticated manner. To illustrate this, we follow Ref. \onlinecite{PhysRevB.80.174401}, and define a local spin operator by
\begin{equation*}
\mathbf{S}(\mathbf{r})=\sum_{ss'}a^\dagger_{s}(\mathbf{r}){\boldmath\text{$\sigma$}}_{ss'}a_{s}(\mathbf{r}),
\end{equation*}
where {\boldmath$\sigma$} is $(\sigma^x,\sigma^y,\sigma^z)$, and $a^\dagger_{\sigma}(\mathbf{r})$ ($a_{\sigma}(\mathbf{r})$) represents local field creation and annihilation operators. An explicit expression for these operators is given in Eq. \eqref{fulla}.
The local magnetization can then be expressed in terms of the spin operator,
\begin{equation}
\mathbf{M}(\mathbf{r})=-\frac{g\mu_B}{V}\sum_{ss'}\langle \mathbf{S(\mathbf{r})}\rangle
\end{equation}
where the average is taken with respect to the dominant ground state obtained from the RG. Here, $g$ is the $g$-factor, $\mu_B$ is the Bohr magneton, and $V$ is the volume of the system. If we expand the local field operator into Bloch waves \eqref{fulla}, we obtain
\begin{equation}\label{SDWM}
\begin{split}
\mathbf{M}(\mathbf{r})=-\frac{g\mu_B}{V}\sum_{ss'}\sum^{c,f}_{A,B}\sum_{K_2K_1}u^\ast_{A,K_2}(\mathbf{r})u_{B,K_1}(\mathbf{r})\\\times e^{-i(K_2-K_1)\cdot\mathbf{r}_{\parallel}}\langle A^\dagger_{K_2,s}{\boldmath\text{$\sigma$}}_{ss'}B_{K_1,s'}\rangle ,
\end{split}
\end{equation}
where the Bloch wave function is $\varphi_{A,K}(\mathbf{r})=e^{-iK\cdot\mathbf{r}_{\parallel}}u_{A,K}(\mathbf{r})$,  $\mathbf{r}=\mathbf{r}_{\perp}+\mathbf{r}_{\parallel}$, $\mathbf{r}_{\perp}$ is the out-of-plane vector,  $\mathbf{r}_{\parallel}$ is the in-plane vector, and $u_{A,K}(\mathbf{r})$ is a periodic function with $\mathbf{r}_{\parallel}\rightarrow \mathbf{r}_{\parallel}+m\mathbf{a}_1$+n$\mathbf{a}_2$, where $m$, $n$ are integers.

Let us also introduce a $SDW$ gap function,
\begin{equation}
\frac{\Delta^{sdw}_{ss'}}{g_1+u_1}=\sum_{{\alpha\alpha'};k}S^z\langle c^\dagger_{\alpha s,k}\tau^x_{\alpha\alpha'}\otimes\sigma^z_{ss'}f_{\alpha' s,k}\rangle.
\end{equation}
If we confine the system to 2D, setting $u_{A,K}(\mathbf{r})=\delta(\mathbf{r}_\perp)$, we  can reduce \eqref{SDWM} to a simpler form,
\begin{equation}
\mathbf{M}(\mathbf{r})=-\frac{2g\mu_B S^z\Delta^{sdw}}{V (g_1+u_1)}\cos(\mathbf{Q}\cdot\mathbf{r}_{\parallel})\delta(\mathbf{r}_\perp).
\end{equation}

Using this formulation, pairing in the $SDW$ channel($c^\dagger_{\mathbf{K}s}(k)\sigma^z_{ss'}f_{\mathbf{K'}s'}(k)$) can be easily identified by this  observable with ordering vector $\mathbf{Q}$. Analogously,  the  $FM$ and $FM'$ pairing channels can be identified.  Similary, for $CDW$, we introduce the local charge density operator,
\begin{equation}
\rho(\mathbf{r})=\frac{1}{V}\sum_{\sigma}\langle a_\sigma^\dagger(\mathbf{r})a_\sigma(\mathbf{r})\rangle.
\end{equation}

From the mean field Hamiltonian (see Appendix \ref{MF}), the trial ground state solutions for $EI$ and $EI'$ are equivalent to Ref. \onlinecite{PhysRev.158.462}. In these excitonic insulating states, the electrons from the conduction band and the holes from the valence band form bound states. 

Furthermore, $FM'$ and $EI'$ break valley symmetry, i.e. time reversal symmetry, because in these ground states, the symmetry exchanging $\mathbf{K}$ and $\mathbf{K'}$ is absent. This can lead to non-trivial insulating states\cite{PhysRevLett.95.226801,PhysRevLett.100.156401,RevModPhys.83.1057}. Since the mean field Hamiltonian in Appendix \ref{MF} does not have a clear `inverted' band gap, we do not conclude that these are quantum spin Hall or quantum anomalous Hall insulator states.

\subsection{Case II: $g_0$, $g_1$, $g_2$ $\ne$ 0}

For non-vanishing values of $g_{0,1,2}$, much of the discussion is similar to the previous section. However, since $g_{0,1,2}$ connect different channels  in the flow rate equations, they do not give simple analytical results that show how the $\Gamma_j$ evolve. Instead, in this more general case  the flow rate equations in \eqref{RGflow} need to be solved numerically. 

\begin{figure}[h]
  \begin{center}
\includegraphics[scale=0.95]{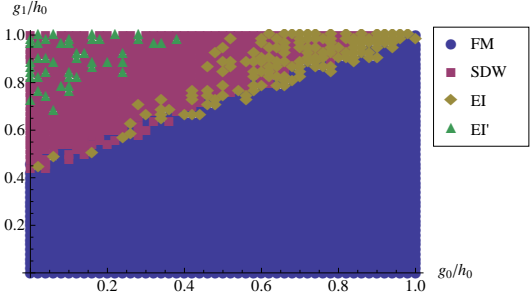}
  \end{center}
   \caption{Phase Diagram for a representative choice of bare couplings $h_0=u_0=v_0=1$, $h_1=v_1=u_1=0.8h_0$,  and $h_2=u_2=v_2=g_2=0.1h_0$. The phase diagram is determined by monitoring which channels divergence first during the RG flow.} \label{Phase}
\end{figure}

Due to the  large parameter space spanned by the possible sets of bare couplings, it is impossible to explore the entire phase diagram. In this section, we only select several regions to scan, illustrating how finite values of $g_{0,1,2}$ affect the results from the previous section. Using the bare values from Fig. \ref{Gammaflow1}, we scan $g_0$ and $g_1$. When $g_{0,1.2}$ is small, we obtain results very similar to the $g_{0,1,2}=0$ case, with $FM$ occupying large regions of the phase diagram. However, when $g_1$ becomes large, we instead obtain the more complicated phase diagram shown in Fig. \ref{Phase}.

\begin{figure}
\begin{center}
\includegraphics[scale=1.1]{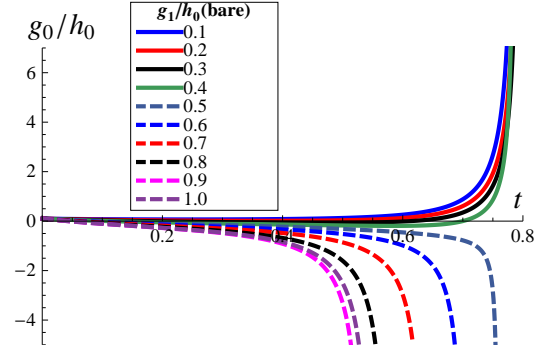}
\end{center}
\caption{The flow of $g_{0}$ as the bare value of $g_{1}$ varies. We set $g_0=g_2=0.1h_0$, and the remaining bare couplings are same as Fig. \ref{Gammaflow1}}\label{g0flow}
\end{figure}

To form an excitonic insulator, $EI$ and $EI'$ order  intricately compete with other instabilities. Without the scattering processes  $g_{0,1,2}$, fine-tuning bare couplings to enhance these instabilities and suppress the others is inevitable. However,  introducing finite $g_{0,1,2}$, the flow of $g_{0,1,2}$ significantly affects this result, which can automatically enhance or suppress the orders. For instance, when $g_1$ is small, the  $g_0$ starts flowing towards more positive values (see Fig. \ref{g0flow}). Consequently, the $FM$ ordering tendency is enhanced (which enhances the divergence of $\Gamma_{FM}$). On the other hand, in Fig. \ref{g0flow}, when $g_1$ becomes large, $g_0$ flows towards increasingly negative values, this suppressing $FM$ order. Because of this suppression, $EI$, $EI'$ and $SDW$  emerge in the large $g_1$ region as shown in Fig. \ref{Phase}.

Furthermore, charge density wave order is  very unlikely to dominate, since $\Gamma_{CDW}(0)=\Gamma_{SDW}(0)-2(v_2+g_2)$, and $g_2$ always grows into the positive regime, as long as only repulsive interactions are considered. We observe that the divergence of spin density wave order is always stronger than charge density wave order. For the same reason, $\Gamma_{FM}(0)=\Gamma_{FM'}(0)-2(u_2+g_2)$, and thus $FM$ order is more favorable than $FM'$.

Similarly, superconducting order is not expected to dominate for small bare values $h_2$ and $g_2$. In order to produce dominant BCS instabilities one needs that $h_2+g_2>h_1+g_1$ or $h_2+g_1>h_1+g_2$ , such that $\Gamma_{SC}(0)$ or $\Gamma_{SC'}(0)$ is positive.

To decide which one is the correct ground state for BLG is beyond the analysis of this paper, because reliable bare coupling constants are in general difficult to obtain. The ultimate answer will require more  experimental input.

\section{Summary}

Summarizing this work, BLG can been modeled by  nearest-neighbor hopping model on a bilayer honeycomb structure with $\gamma_3\simeq0$. A general form of interactions between electrons can be accounted for without including long-range Coulomb  interactions. Due to the trivial topology of the Fermi surface, the RG tree level analysis eliminates  irrelevant channels and greatly simplifies the interacting terms in this model. The RG flow rate equation can be calculated up to one-loop level in the weak coupling limit. 

Instabilities are inevitable if the inter-band and inter-valley interactions are nonzero. We have investigated each instability, and related them to ordered ground states. Specifically, we have found competing ferromagnetic ($FM$, $FM'$) , spin density wave ($SDW$), excitonic insulator ($EI$, $EI'$),  charge density wave ($CDW$), and superconducting ($SC$, $SC'$) ground states in this model. 
Except $SC$ and $SC'$, all the ground states are insulating. Furthermore, valley symmetry breaking is found in the $FM'$ and $EI'$  channels.

Since the free system with quadratic dispersion is not stable, any small perturbation can drive this point toward divergence. Due to particle-hole symmetry and perfect nesting of the two Fermi points $\mathbf{K}$, $\mathbf{K'}$, excitonic instabilities are expected to arise, which has been previously pointed out in Ref. \onlinecite{PhysRevLett.104.156803, PhysRevB.82.115431}.   

\section{Discussion}\label{discussion}

Projecting out the orbital field operators $a_{1K\sigma}$ and $a_{2K\sigma}$ in \eqref{Eigenstate} and taking the spatial continuum limit, the non-interacting low energy effective model of BLG can be approximated by a massive chiral fermion model\cite{PhysRevLett.96.086805}. By symmetry arguments,  all  possible two-body interactions can be obtained.\cite{Vafek:2010kx,Zhang:2010vn,PhysRevB.82.201408} Under this limit, this model exhibit a very rich and exotic low energy phenomenology because of the newly emerging valley and pseudospin degree of freedom. 

In this paper, we utilize the band representation point of view which are not necessary to impose continuum limit. Projecting out the gapped bands, BLG is effectively viewed as a conventional two-band model and all its interactions can be immediately obtained according to the band index. In this approach, the instabilities of BLG is clearly interpreted as the peculiar nature of the FS and the low energy physics exhibit rich excitonic orders.

In addition, since we are performing RG transformation in band representation, the `which-layer' \cite{PhysRevLett.104.156803} or pseudospin\cite{PhysRevB.77.041407} symmetry breaking is not obvious, and any `which-layer' order cannot be straight-forwardly observed from the order parameters. As discussed in the previous section, the order parameters are pairings between the $c$  and $f$ bands, thus giving rise to a complicated relation for local magnetization and charge density. To extract layer order, it would be necessary to know the appropriate Bloch wave function or its Wannier representation. Comparing to the approach from Ref. \onlinecite{Vafek:2010kx,Zhang:2010vn,PhysRevB.82.201408,2011arXiv1111.2076T,2012arXiv1206.0288C}, this is one of the drawbacks of our approach.

Furthermore, the ground states in this paper have been classified according to their band index, and the  pseudospin index is implicitly contained in \eqref{Eigenstate}. Therefore, the pseudospin symmetry breaking is not made explicit in our model. This leads to a different physical interpretation of the ordered states. Because of this reason, not all the of the possible competing ground states found in Ref. \onlinecite{2012arXiv1206.0288C} can be obtained by in our study.

A recent functional renormalization group (fRG) study\cite{2011arXiv1112.5038S} has demonstrated the advantage of retaining all the lattice structure, and integrating out energy modes without ambiguities. Furthermore, their approach takes into account the complication of angular dependence in the interactions. Their study has shown an interesting ``three-sublattice CDW instability''. This instability is quickly disappears as the on-site interaction becomes dominant. Since their model Hamiltonian is different from the one studied in this paper, a direct comparison with their results is not straightforward. One of the big discrepancies is the predominant $FM$ instability observed in our approach. This may arise because exchange interactions\cite{PhysRevB.85.075438} were not explicitly considered in Ref. \onlinecite{2011arXiv1112.5038S}. 

The results in this paper are valid only of the  one-loop level. Typically, higher-loop contributions can be neglected by invoking $1/N$ arguments.\cite{Shankar:1994uq} However, this type of argument is not very strong for this model, because the Fermi surface contains just two points, resulting in $N=4$ only. Also, the results presented here only apply for the weak coupling limit. Any strong enough coupling to break down the perturbative expansion will invalidate the preceding discussion. In the strong coupling limit, results from the tree level analysis cannot be trusted, and using the same effective action as in \eqref{SintF} will not guarantee correct results. When the order of the tree  and loop diagrams is comparable, new effective models and non-perturbative  approaches may be needed. 

In parameter space, the point with non-zero trigonal warping, i.e. the free part of the action is non-analytic. Due to this reason, perturbative RG may not work properly. In particular, the scaling rule of the fermonic field cannot be defined, and one does not know whether this point is a Gaussian fixed-point or not. Therefore, we enforce using $\gamma_3=0$ as the fixed-point to define the scaling rule, and always treat the trigonal warping term as a quadratic perturbation.

Furthermore, the instabilities in this paper are driven by perfect particle-hole nesting. The presence of disorder can destroy this symmetry and thus change the phase diagram. Doping away from the charge neutrality point, the FS becomes a line rather than a few points. In this case, the scaling properties of the fermionic field are different, and the conclusions of our approach are no longer valid. As show in a recent study of doped monolayer graphene,\cite{PhysRevB.86.020507} functional RG is a promising alternative method to study the BLG doping problem. In the doped case, functional RG may superior to our approach, since the shape of FS evolves nontrivially under doping.

The results of our model are consistent with recent current transport spectroscopic experiments.\cite{VelascoJ.:2012fk,PhysRevLett.108.076602} Specifically, a magnetic field dependent gap is expected in a ground state with excitonic order. \cite{PhysRev.170.816} Therefore, magnetically order ground states are not a necessary  condition to exhibit this property. 
To make  connections with experiments, the physical properties of the ground states discussed in this paper need to be analyzed, in particular how these ground states respond to external perturbations, especially to currents. Also in the experiments, considering the effects of disorder and boundaries is important.

We would like to thank Tameem Albash, Rahul Nandkishore, Ronny Thomale, Hubert Saleur, Oscar Vafek, and Lorenzo Campos Venuti for useful discussions, and acknowledge financial support by the Department of 
Energy under grant DE-FG02-05ER46240.

\appendix
\section{Coupling Constants and the Interacting Hamiltonian}\label{Ainteractions}

In this appendix, we show how the interacting Hamiltonian for the BLG model is constructed. To accomplish this, we approximate the Bloch wave function by using the $\pi_z$ orbital $\phi(\mathbf{r})=\sqrt{\xi^5/\pi}ze^{-\xi r} $ with $\xi=1.72 a_0^{-1}$,\cite{PhysRevB.85.075438}
\begin{equation}\label{Bloch}
\begin{split}
&\varphi_{IK}(\mathbf{r})= \\ 
&\quad\sum^{MN}_{mn;i}\frac{C^I_{iK}}{\sqrt{N_{uc}}}e^{iK\cdot \mathbf{R}_{mn}} \phi(\mathbf{r}-\mathbf{R}_{mn}-\mathbf{t}_i),
\end{split}
\end{equation}
where $N_{uc}$ is the number of unit cells, $a_0$ is the Bohr radius, $I=c,f$ bands, $\mathbf{R}_{mn}=m \mathbf{a}_1+ n \mathbf{a}_2$ denote the center of the unit cell,  and $t_m$ are the basis in the unit cell. $K$ is the in-plane crystal momentum of BLG in the first Brillouin zone. $i=1$ represents the site $b_2$ with $\mathbf{t}_1=(0,0,-c)$, $i=2$ represents the site $a_2$ with  $\mathbf{t}_2=(2a,0,-c)$, $i=3$ represents the site $a_1$ with $\mathbf{t}_3=(2a,0,0)$, and $i=4$ represents the site $b_1$ with $\mathbf{t}_4=(a,0,0)$ (see Fig. \ref{BLG}). $a$ is the lattice spacing $1.57 \AA$, and $c$ is the layer separation $3.35 \AA$.

Graphene can be considered as a 2D material, but the electrons still live in 3D real space. In order to obtain correct interaction terms between electrons, we  start from the original Hamiltonian\cite{SimonQFT}(Born-Oppenheimer approximation is used), which describes the electrons with Coulomb interaction in real space representation,
\begin{equation*}\label{Hfull}\begin{split}
&H_{full}=\int d^3\mathbf{r} a^\dagger_{\sigma}(\mathbf{r})[\frac{-\hbar^2}{2m}\nabla^2+V_{ext}(\mathbf{r})]a_{\sigma}(\mathbf{r}) +\\
&\frac{1}{2}\int d^3\mathbf{r}d^3\mathbf{r'}a^\dagger_{\sigma}(\mathbf{r})a^\dagger_{\sigma'}(\mathbf{r'})V_{int}(\mathbf{r}-\mathbf{r'})a_{\sigma'}(\mathbf{r'})a_{\sigma}(\mathbf{r}),
\end{split}
\end{equation*}
where $a^\dagger_{\sigma}(\mathbf{r})$ ($a_{\sigma}(\mathbf{r})$) are local field operator which create (annihilate) an electron at $\mathbf{r}$. $V_{ext}$ is the potential produced by the ions. $V_{int}$ is the Coulomb potential between electrons at $\mathbf{r}$ and $\mathbf{r'}$.

Now we approximate (the gapped bands are neglected) the full operator $a_\sigma(\mathbf{r})$ by expanding it into Bloch waves from (\ref{Bloch}). Then  we have
\begin{equation}\label{fulla}
\begin{split}
a^\dagger_{\sigma}(\mathbf{r})&\simeq\sum_{K}[\varphi^\ast_{c,K}(\mathbf{r}) c^\dagger_{\sigma K}+\varphi^\ast_{f,K}(\mathbf{r}) f^\dagger_{\sigma K}],\\
a_{\sigma}(\mathbf{r})&\simeq\sum_{K}[\varphi_{c,K}(\mathbf{r}) c^\dagger_{\sigma K}+\varphi_{f,K}(\mathbf{r}) f^\dagger_{\sigma K}].
\end{split}
\end{equation}
By using
\begin{equation*}
[\frac{-\hbar^2\nabla^2}{2m}+V_{ext}(\mathbf{r})]\varphi_{c,f;K}(\mathbf{r})=\epsilon_{c,f}(K)\varphi_{c,f;K}(\mathbf{r})
\end{equation*}
and substituting the above equation into $H_{full}$,  one can easily obtain $H_0$ in (\ref{H0}) and $H_{int}$ in (\ref{Hint}) and \eqref{MixedInt}. The coupling constant is determined by 
\begin{equation}\label{U4321}
\begin{split}
\mathcal{U}(K_3K_4K_2K_1)=\int d^3\mathbf{r}d^3\mathbf{r'}V_{int}(\mathbf{r}-\mathbf{r'})\\
\times\varphi^\ast_{I_3K_3}(\mathbf{r})\varphi^\ast_{I_4K_4}(\mathbf{r'})\varphi_{I_2K_2}(\mathbf{r'})\varphi_{I_1K_1}(\mathbf{r}).
\end{split}
\end{equation}
By substituting (\ref{Bloch}) into (\ref{U4321}), one can also verify that the valley and particle-hole symmetries still hold, yielding only six independent coupling constants. Note that  projectinging out the gapped bands introduces a hard cutoff, further modifying $\mathcal{U}$, which should behave like the original long-range Coulomb interaction\cite{Shankar:1994uq}.

\subsection{Interactions in the BLG Hamiltonian}

From Eqs. \eqref{U4321} and \eqref{Hint} we find ten inequivalent interaction terms, which are not ruled out by  symmetry and conservation laws,\cite{PhysRevB.84.205445}
\begin{eqnarray}
\mathcal{U}_4(K_3K_4K_2K_1)f^{\dagger}_{\sigma K_3}c^{\dagger}_{\sigma'K_4}c_{\sigma'K_2}c_{\sigma K_1}\nonumber\\
\mathcal{U}_4(K_3K_4K_2K_1)c^{\dagger}_{\sigma K_3}f^{\dagger}_{\sigma'K_4}f_{\sigma'K_2}f_{\sigma K_1}\nonumber\\
\mathcal{U}_5(K_3K_4K_2K_1)c^{\dagger}_{\sigma K_3}c^{\dagger}_{\sigma^\prime K_4}c_{\sigma'K_2}f_{\sigma K_1}\nonumber\\
\mathcal{U}_5(K_3K_4K_2K_1)f^{\dagger}_{\sigma K_3}f^{\dagger}_{\sigma'K_4}f_{\sigma'K_2}c_{\sigma K_1}\label{MixedInt}
\end{eqnarray}

Of these, $\mathcal{U}_4$ and $\mathcal{U}_5$ are irrelevant under RG tree level, because they  vanish at the FS. In the following, we will prove $\mathcal{U}_4(\mathbf{K},\mathbf{K},\mathbf{K},\mathbf{K})=0$, and the proof for different valley combination and $\mathcal{U}_5$ is similar.

Using \eqref{U4321}, we have
\begin{equation}
\begin{split}
\mathcal{U}_4(\mathbf{K},\mathbf{K},\mathbf{K},\mathbf{K})=\int d^3\mathbf{x}d^3\mathbf{x'}\varphi^\ast_{v\mathbf{K}}(\mathbf{x})\times\\
\varphi^\ast_{c\mathbf{K}}(\mathbf{x'})V_{int}(\mathbf{x}-\mathbf{x'})\varphi_{c\mathbf{K}}(\mathbf{x})\varphi_{c\mathbf{K}}(\mathbf{x'})
\end{split}
\end{equation}
Note that $C^v_{K}=\frac{1}{\sqrt2}(1,0,0,1)$ and $C^c_{K}=\frac{1}{\sqrt2}(1,0,0,-1)$ at $K=\mathbf{K},\mathbf{K'}$. Now we use Eq.  \eqref{Bloch} to write out the Bloch wave function explicitly,
\begin{widetext}
\begin{equation}\label{U3}
\begin{split}
\mathcal{U}_4(\mathbf{K},\mathbf{K},\mathbf{K},\mathbf{K})=\alpha \sum_{m_1n_1}\sum_{m'_1n'_1}\sum_{m_2n_2}\sum_{m'_2n'_2}\int d^3\mathbf{x}d^3\mathbf{x'}
e^{i\mathbf{K}\cdot(\mathbf{R}_{m_1n_1}+\mathbf{R}_{m'_1n'_1}-\mathbf{R}_{m_2n_2}-\mathbf{R}_{m'_2n'_2})}\times\\
V_{int}(\mathbf{x}-\mathbf{x'})\Phi_v(\mathbf{x}-\mathbf{R}_{m_1n_1})\Phi_c(\mathbf{x'}-\mathbf{R}_{m'_1n'_1})\Phi_c(\mathbf{x'}-\mathbf{R}_{m'_2n'_2})\Phi_c(\mathbf{x}-\mathbf{R}_{m_2n_2}),
\end{split}
\end{equation}
\end{widetext}
where $\Phi_{v}(\mathbf{x})=\frac{1}{\sqrt2}[\phi(\mathbf{x}-\mathbf{t}_1)-\phi(\mathbf{x}-\mathbf{t}_4)]$, $\Phi_{c}(\mathbf{x})=\frac{1}{\sqrt2}[\phi(\mathbf{x-\mathbf{t}_1})+\phi(\mathbf{x}-\mathbf{t}_4)]$, and $\alpha$ is some constant. Applying changes of variables, $\mathbf{x}\rightarrow\mathbf{x}+(\mathbf{t}_1+\mathbf{t}_4)$ and $\mathbf{x'}\rightarrow\mathbf{x'}+(\mathbf{t}_1+\mathbf{t}_4)$, and focusing on the $\Phi_v(\mathbf{x}-\mathbf{R}_{m_1n_1})\Phi_c(\mathbf{x}-\mathbf{R}_{m_2n_2})$ product term in the integrand. 
\begin{widetext}
\begin{equation}
\begin{split}
&\Phi_v(\mathbf{x}-\mathbf{R}_{m_1n_1})\Phi_c(\mathbf{x}-\mathbf{R}_{m_2n_2})\\
&=\frac{1}{2}(\phi(\mathbf{x}-\mathbf{R}_{m_1n_1}-\mathbf{t}_1)-\phi(\mathbf{x}-\mathbf{R}_{m_2n_2}-\mathbf{t}_4))(\phi(\mathbf{x}-\mathbf{R}_{m_1n_1}-\mathbf{t}_1)+\phi(\mathbf{x}-\mathbf{R}_{m_2n_2}-\mathbf{t}_4))\\
&\rightarrow\frac{1}{2}(\phi(\mathbf{x}-\mathbf{R}_{m_1n_1}+\mathbf{t}_4)-\phi(\mathbf{x}-\mathbf{R}_{m_2n_2}+\mathbf{t}_1))(\phi(\mathbf{x}-\mathbf{R}_{m_1n_1}+\mathbf{t}_4)+\phi(\mathbf{x}-\mathbf{R}_{m_2n_2}+\mathbf{t}_1)).
\end{split}
\end{equation}
Next, we  perform changes of variables for $m$ and $n$, $m\rightarrow M-m$ and $n\rightarrow N-n$. Therefore, $\mathbf{R}_{mn}\rightarrow\mathbf{R}_{MN}-\mathbf{R}_{mn}$ and this does not affect anything but,
\begin{equation*}
\begin{split}
\Phi_v(\mathbf{x}-\mathbf{R}_{m_1n_1})\Phi_c(\mathbf{x}-\mathbf{R}_{m_2n_2})\rightarrow&
\frac{1}{2}(\phi(\mathbf{x}-\mathbf{R}_{MN}+\mathbf{R}_{m_1n_1}+\mathbf{t}_4)-\phi(\mathbf{x}-\mathbf{R}_{MN}+\mathbf{R}_{m_2n_2}+\mathbf{t}_1))\\ \times &(\phi(\mathbf{x}-\mathbf{R}_{MN}+\mathbf{R}_{m_1n_1}+\mathbf{t}_4)+\phi(\mathbf{x}-\mathbf{R}_{MN}+\mathbf{R}_{m_2n_2}+\mathbf{t}_1))
\end{split}
\end{equation*}
\end{widetext}
$\mathbf{R}_{MN}$ can be removed by performing changes of variables $\mathbf{x}\rightarrow-\mathbf{x}+\mathbf{R}_{MN}$ and $\mathbf{x'}\rightarrow-\mathbf{x'}+\mathbf{R}_{MN}$. Notice that the $\pi_z$ orbital satisfies $\phi(-\mathbf{x})=-\phi(\mathbf{x})$. Therefore, we obtain exactly the same expression as in \eqref{U3}, except a minus sign. This means $\mathcal{U}_4$ must be vanish at FS. Similarly, $\mathcal{U}_5=0$ for the same reason. 

From the result of the RG tree level, all couplings with small $k$ dependence are irrelevant. The first leading non-vanishing term in $\mathcal{U}_4$, $\mathcal{U}_5$ is $\mathcal{O}(k)$.  Therefore, the interactions in \eqref{MixedInt} are irrelevant.

\subsection{Coupling Constant Expansion}

Here we show how to expand the coupling constants around  $\mathbf{K}$ and $\mathbf{K'}$. First we perform a unitary transformation by changing from Bloch wave to Wannier representation. The Wannier function is defined as 
\begin{equation}\label{Wannier2Bloch}
w_{I,mn}(\mathbf{r})=\sum^{B.Z.}_{K}\frac{e^{-iK\cdot \mathbf{R}_{mn}}}{\sqrt{N_{uc}}}\varphi_{IK}(\mathbf{r}),
\end{equation}
and
\begin{equation}\label{Bloch2Wannier}
\varphi_{IK}(\mathbf{r})=\sum^{M,N}_{m,n=0}\frac{e^{iK\cdot \mathbf{R}_{mn}}}{\sqrt{N_{uc}}}w_{I,mn}(\mathbf{r}),
\end{equation}
where $N_{uc}$ is the total number of unit cells. Also from \eqref{Wannier2Bloch} and \eqref{Bloch2Wannier}, we can derive the identity
\begin{equation}\label{delta}
\sum_{mn}e^{iK\cdot\mathbf{R}_{mn}}=(2\pi)^2N_{uc}\bar{\delta}^2(K)
\end{equation}
Note that, $e^{iK\cdot\mathbf{R}_{mn}}=e^{iK\cdot\mathbf{R}_{mn}+\mathbf{G}\cdot\mathbf{R}_{mn}}$, where $\mathbf{G}$ is a reciprocal lattice vector. Therefore  $\bar{\delta}(K)$ in \eqref{delta} is not exactly the Dirac delta function, but equal to a delta function up to a reciprocal lattice vector.

Now, using \eqref{delta} and (\ref{U4321}), we obtain
\begin{widetext}
\begin{equation*}
\begin{split}
\mathcal{U}(K_3K_4K_2K_1)=&\frac{1}{N^2_{uc}}\sum_{n_3m_3}\sum_{n_4m_4}\sum_{n_2m_2}\sum_{n_1m_1}e^{-iK_3\cdot\mathbf{R}_{m_3n_3}}e^{-iK_4\cdot\mathbf{R}_{m_4n_4}}e^{iK_2\cdot\mathbf{R}_{m_2n_2}}e^{iK_1\cdot\mathbf{R}_{m_1n_1}}\\
&\int d^3\mathbf{x}d^3\mathbf{x^\prime}w^\ast_{I_3,m_3n_3}(\mathbf{x})w^\ast_{I_4,m_4n_4}(\mathbf{x^\prime})V'_{int}(\mathbf{x}-\mathbf{x^\prime})w_{I_2,m_2n_2}(\mathbf{x'})w_{I_1,m_1n_1}(\mathbf{x})
\end{split}
\end{equation*}
\begin{equation*}
\begin{split}
\mathcal{U}(K_3K_4K_2K_1)=&\frac{1}{N^2_{nc}}\sum_{n_3m_3}\sum_{n_4m_4}\sum_{n_2m_2}\sum_{n_1m_1}e^{-i(K_4+K_3-K_2-K_1)\cdot\mathbf{R}_{m_4n_4}}\times\\
&e^{-iK_3\cdot(\mathbf{R}_{m_3n_3}-\mathbf{R}_{m_4n_4})}e^{iK_2\cdot(\mathbf{R}_{m_2n_2}-\mathbf{R}_{m_4n_4})}e^{iK_1\cdot(\mathbf{R}_{m_1n_1}-\mathbf{R}_{m_4n_4})}\times\\
&\int d^3\mathbf{x}d^3\mathbf{x^\prime}w^\ast_{I_3,m_3n_3}(\mathbf{x})w^\ast_{I_4,m_4n_4}(\mathbf{x^\prime})V_{int}(\mathbf{x}-\mathbf{x^\prime})w_{I_2,m_2n_2}(\mathbf{x'})w_{I_1,m_1n_1}(\mathbf{x})
\end{split}
\end{equation*}
by changes of variables $\mathbf{x}\rightarrow\mathbf{x}+\mathbf{R}_{m_4n_4}$, $\mathbf{x'}\rightarrow\mathbf{x'}+\mathbf{R}_{m_4n_4}$ and using (\ref{delta}),
 \begin{equation}\label{expansion}
\begin{split}
\mathcal{U}(K_4K_3K_2K_1)=&\frac{(2\pi)^2}{N_{uc}}\bar{\delta}^2(K_4+K_3-K_2-K_1)\sum_{n'_3m'_3}\sum_{n'_2m'_2}\sum_{n'_1m'_1}
e^{-iK_3\cdot\mathbf{R}_{m'_3n'_3}}e^{iK_2\cdot\mathbf{R}_{m'_2n'_2}}e^{iK_1\cdot\mathbf{R}_{m'_1n'_1}}\times\\
&\int d^3\mathbf{x}d^3\mathbf{x^\prime}w^\ast_{I_3,00}(\mathbf{x})w^\ast_{I_4,m'_4n'_4}(\mathbf{x^\prime})V_{int}(\mathbf{x}-\mathbf{x'})w_{I_2,m'_2n'_2}(\mathbf{x'})w_{I_1,m'_1n'_1}(\mathbf{x})
\end{split}
\end{equation}
\end{widetext}

The coupling constant expansion can be achieved  by expanding $e^{-iK_3\cdot\mathbf{R}_{m'_3n'_3}}$, $e^{iK_2\cdot\mathbf{R}_{m'_2n'_2}}$, $e^{iK_1\cdot\mathbf{R}_{m'_1n'_1}}$ in Eq. \eqref{expansion}.
$\bar{\delta}(K_4+K_3-K_2-K_1)$ means that  momentum is conserved up to a reciprocal lattice vector. This allows Umklapp processes. Momentum conservation emerges because of the in-plane lattice translational invariance in the system.

\section{Renormalization Group and Tree Level Analysis}\label{RGTree}

This section summarizes the details needed for the perturbative renormalization group analysis.

\subsection{Action of the Model Hamiltonian}

The RG transformation is performed using the path integral formalism. Therefore, the model Hamiltonian from Section 2 should be rewritten into action form. $S=\int d\tau \mathcal{L}=S_0+S_{int}$, where $S_0$ is the free action and $S_{int}$ contains the interaction terms. The derivation is tedious, and one needs to introduce coherent states of the creation and annihilation field operators\cite{Shankar:1994uq}. However, the result is simple, which can be achieved by replacing the field creation and annihilation operator by Grassmann fields. Namely, $c^\dagger_{\sigma}\rightarrow\bar{\psi}_\sigma$, $c_\sigma\rightarrow\psi_\sigma$ and $f^\dagger_{\sigma}\rightarrow\bar{\chi}_\sigma$, $f\rightarrow\chi_\sigma$. Therefore, the $S_0$ and  $S_{int}$ can be written as 
\begin{widetext}
\begin{equation*}
\begin{split}
S_0&=\int_{-\infty}^{\infty}\frac{d\omega}{2 \pi}\int_{\substack{|K-\mathbf{K}|\leq\Lambda \\ |K-\mathbf{K^\prime}|\leq\Lambda }}\frac{d^2K}{(2\pi)^2} \bar{\psi}_{\sigma}(K,\omega)(i\omega-\epsilon_{c}(K))\psi_{\sigma}(K,\omega)+\bar{\chi}_{\sigma}(K,\omega)(i\omega-\epsilon_{v}(K))\chi_{\sigma}(K,\omega)
\end{split}
\end{equation*}
\begin{equation}\label{Sint}
\begin{split}
S_{int}=\frac{1}{2}\bigl[\prod_{i=1}^4\int_{-\infty}^{\infty}\frac{d\omega_i}{2\pi}&\int_{\substack{|K_i-\mathbf{K}|\leq\Lambda \\ |K_i-\mathbf{K^\prime}|\leq\Lambda }}\frac{d^2K_i}{(2\pi)^2}\bigr]2\pi\delta(\omega_1+\omega_2-\omega_3-\omega_4)\times\\ 
&\bigl\{
\;\;\,\mathcal{U}_0(K_3K_2K_2K_1)\bar{\psi}_{\sigma}(K_3,\omega_3)\bar{\psi}_{\sigma^\prime}(K_4,\omega_4)\psi_{\sigma^\prime}(K_2,\omega_2)\psi_{\sigma}(K_1,\omega_1)\\
&+ \mathcal{U}_1(K_3K_4K_2K_1)\bar{\psi}_{\sigma}(K_3,\omega_3)\bar{\psi}_{\sigma^\prime}(K_4,\omega_4)\chi_{\sigma^\prime}(K_2,\omega_2)\chi_{\sigma}(K_1,\omega_1)\\
&+ \mathcal{U}_2(K_3K_4K_2K_1)\bar{\psi}_{\sigma}(K_3,\omega_3)\bar{\chi}_{\sigma^\prime}(K_4,\omega_4)\chi_{\sigma^\prime}(K_2,\omega_2)\psi_{\sigma}(K_1,\omega_1)\\
&+ \mathcal{U}_3(K_3K_4K_2K_1)\bar{\chi}_{\sigma}(K_3,\omega_3)\bar{\psi}_{\sigma^\prime}(K_4,\omega_4)\chi_{\sigma^\prime}(K_2,\omega_2)\psi_{\sigma}(K_1,\omega_1)
\bigr\}\\
&+ [\text{exchange } (\psi\leftrightarrow\chi)]
\end{split}
\end{equation}
\end{widetext}

Now we can perform the RG analysis. First, $S_0$ is chosen to be the fixed point in the theory. This choice will determine the scaling properties of $\omega$ and $\psi$, which will  be discussed in the following section.

\subsection{Scaling Properties and Effective Action at the Tree Level}

Since we interested in low energy limit, only the energy modes in the vicinity of  Fermi points are considered (see Fig. \ref{1BZ}). Expanding $\epsilon_{c,f}(K)$ around $\mathbf{K}$ and $\mathbf{K'}$, and combining with the results from  (\ref{dispersion}),
\begin{equation}
\begin{split}
S_0&
=\sum_{\alpha=\mathbf{K},\mathbf{K^\prime}}\int_{-\infty}^{\infty}\frac{d\omega}{2 \pi}\int_{|k|\leq\Lambda}\frac{d^2k}{(2\pi)^2} \times\\
&
\begin{bmatrix}\bar{\psi}_{\alpha\sigma}(k,\omega)(i\omega-\frac{v_F^2}{\gamma_1}k^2)\psi_{\alpha\sigma}(k,\omega)\\
+\bar{\chi}_{\alpha\sigma}(k,\omega)(i\omega+\frac{v_F^2}{\gamma_1}k^2)\chi_{\alpha\sigma}(k,\omega)
\end{bmatrix}.
\end{split}
\end{equation}

We introduce a short hand notation $\bar{\psi}_{\alpha\sigma}(k,\omega)=\bar{\psi}_{\sigma}(\alpha+k,\omega)$, $\psi_{\alpha\sigma}(k,\omega)=\psi_{\sigma}(\alpha+k,\omega)$, $\bar{\chi}_{\alpha\sigma}(k,\omega)=\bar{\chi}_{\sigma}(\alpha+k,\omega)$, $\chi_{\alpha\sigma}(k,\omega)=\chi_{\sigma}(\alpha+k,\omega)$, and $\alpha=\mathbf{K},\mathbf{K^\prime}$ is known as the `valley' degree of freedom.Valley index is similar to L (left) and R (right) index in the one dimensional case.\cite{Shankar:1994uq}

The RG transformation is simply integrating out the high energy modes of $\bar{\psi}_{\alpha\sigma}(k,\omega)$, $\psi_{\alpha\sigma}(k,\omega)$ and $\bar{\chi}_{\alpha\sigma}(k,\omega)$, $\chi_{\alpha\sigma}(k,\omega)$ which lie within the thin shell, $\Lambda$, in Figure \ref{1BZ}, and considering how these modes affect the low energy theory. After integrating out, only $|k'|\leq \Lambda-d\Lambda=\Lambda/s$ modes remain in the theory. In order to evaluate what has changed from the original theory, $k^\prime$ must be rescaled ($k'= sk$) back to the original phase space such that $|k|\leq\Lambda$.
Since $S_0$ is the fixed point, this requires that $\omega$, $\psi_{\alpha,\sigma}(k,\omega)$ and $\chi_{\alpha,\sigma}(k,\omega)$ must be rescaled,
\begin{eqnarray*}
\omega^\prime&=&s^2\omega,\\
\psi^\prime_{\alpha\sigma}(k',\omega^\prime)&=&s^{-3}\psi_{\alpha\sigma}(k'/s,\omega),\\
\chi^\prime_{\alpha\sigma}(k',\omega^\prime)&=&s^{-3}\chi_{\alpha\sigma}(k'/s,\omega).
\end{eqnarray*}

With this scaling relation, one can now ask how the coupling constants $\mathcal{U}_0$, $\mathcal{U}_1$, $\mathcal{U}_2$, $\mathcal{U}_3$, $\mathcal{U}_4$ and $\mathcal{U}_5$ scale under the RG transformation. Again, we use Eq. \eqref{expansion} to expand couplings around $\mathbf{K}$ and $\mathbf{K^\prime}$. By enforcing momentum conservation, only the constant term in the expansion do not renormalize to zero (marginal under tree level).

Note that $S_{int}$ remains unchanged when $K_4\leftrightarrow K_3$ and $K_2\leftrightarrow K_1$ simultaneously. In addition, using time reversal symmetry (valley symmetry) in the model, hence exchanging $\mathbf{K}\leftrightarrow\mathbf{K'}$ in (Table \ref{coupling}) will not produce another set of independent coupling constants. Thus we obtain $S_{int}$ at the tree level,
\begin{widetext}
\begin{equation}\label{SintF}
\begin{split}
S_{int}=&\frac{1}{2}\bigl[\prod_{i=1}^4\int_{-\infty}^{\infty} \frac{d\omega_i}{2\pi}\int_{|k_i|\leq\Lambda}\frac{d^2k_i}{(2\pi)^2}\bigr](2\pi)^2\bar{\delta}^2(k_1+k_2-k_3-k_4)2\pi\delta(\omega_1+\omega_2-\omega_3-\omega_4)\times\bigl\{ \\
& \quad \begin{bmatrix}
 h_0\bar{\psi}_{\mathbf{K}\sigma}(k_3,\omega_3)\bar{\psi}_{\mathbf{K}\sigma'}(k_4,\omega_4)\psi_{\mathbf{K}\sigma'}(k_2,\omega_2)\psi_{\mathbf{K}\sigma}(k_1,\omega_1)&\\
+h_1\bar{\psi}_{\mathbf{K}\sigma}(k_3,\omega_3)\bar{\psi}_{\mathbf{K'}\sigma'}(k_4,\omega_4)\psi_{\mathbf{K'}\sigma'}(k_2,\omega_2)\psi_{\mathbf{K}\sigma}(k_1,\omega_1)&+\text{exchange } (\mathbf{K}\leftrightarrow\mathbf{K'})\\
 +h_2\bar{\psi}_{\mathbf{K'}\sigma}(k_3,\omega_3)\bar{\psi}_{\mathbf{K}\sigma'}(k_4,\omega_4)\psi_{\mathbf{K'}\sigma^\prime}(k_2,\omega_2)\psi_{\mathbf{K}\sigma}(k_1,\omega_1)& 
\end{bmatrix}\\
&+\begin{bmatrix}
 g_0\bar{\psi}_{\mathbf{K}\sigma}(k_3,\omega_3)\bar{\psi}_{\mathbf{K}\sigma'}(k_4,\omega_4)\chi_{\mathbf{K}\sigma'}(k_2,\omega_2)\chi_{\mathbf{K}\sigma}(k_1,\omega_1)&\\
+g_1\bar{\psi}_{\mathbf{K}\sigma}(k_3,\omega_3)\bar{\psi}_{\mathbf{K'}\sigma'}(k_4,\omega_4)\chi_{\mathbf{K'}\sigma'}(k_2,\omega_2)\chi_{\mathbf{K}\sigma}(k_1,\omega_1)&+\text{exchange } (\mathbf{K}\leftrightarrow\mathbf{K'})\\
 +g_2\bar{\psi}_{\mathbf{K'}\sigma}(k_3,\omega_3)\bar{\psi}_{\mathbf{K}\sigma'}(k_4,\omega_4)\chi_{\mathbf{K'}\sigma'}(k_2,\omega_2)\chi_{\mathbf{K}\sigma}(k_1,\omega_1)&
\end{bmatrix}\\
&+\begin{bmatrix}
u_0\bar{\psi}_{\mathbf{K}\sigma}(k_3,\omega_3)\bar{\chi}_{\mathbf{K}\sigma'}(k_4,\omega_4)\chi_{\mathbf{K}\sigma'}(k_2,\omega_2)\psi_{\mathbf{K}\sigma}(k_1,\omega_1)&\\
+u_1\bar{\psi}_{\mathbf{K}\sigma}(k_3,\omega_3)\bar{\chi}_{\mathbf{K'}\sigma'}(k_4,\omega_4)\chi_{\mathbf{K'}\sigma'}(k_2,\omega_2)\psi_{\mathbf{K}\sigma}(k_1,\omega_1)&+\text{exchange } (\mathbf{K}\leftrightarrow\mathbf{K'})\\
+u_2\bar{\psi}_{\mathbf{K'}\sigma}(k_3,\omega_3)\bar{\chi}_{\mathbf{K}\sigma'}(k_4,\omega_4)\chi_{\mathbf{K'}\sigma'}(k_2,\omega_2)\psi_{\mathbf{K}\sigma}(k_1,\omega_1)&
\end{bmatrix}\\
&+\begin{bmatrix}
v_0\bar{\chi}_{\mathbf{K}\sigma}(k_3,\omega_3)\bar{\psi}_{\mathbf{K}\sigma'}(k_4,\omega_4)\chi_{\mathbf{K}\sigma'}(k_2,\omega_2)\psi_{\mathbf{K}\sigma}(k_1,\omega_1)&\\
+v_1\bar{\chi}_{\mathbf{K}\sigma}(k_3,\omega_3)\bar{\psi}_{\mathbf{K'}\sigma'}(k_4,\omega_4)\chi_{\mathbf{K'}\sigma'}(k_2,\omega_2)\psi_{\mathbf{K}\sigma}(k_1,\omega_1)&+\text{exchange } (\mathbf{K}\leftrightarrow\mathbf{K'}) \\
+v_2\bar{\chi}_{\mathbf{K'}\sigma}(k_3,\omega_3)\bar{\psi}_{\mathbf{K}\sigma'}(k_4,\omega_4)\chi_{\mathbf{K'}\sigma'}(k_2,\omega_2)\psi_{\mathbf{K}\sigma}(k_1,\omega_1)& 
\end{bmatrix}\\
&+[\text{exchange }  (\psi\leftrightarrow\chi) ]\;\bigr\}.
\end{split}
\end{equation}
\end{widetext}

\section{Mean Field Analysis of the Ground States}\label{MF}

In this section, we summarize the mean field analysis for the ground states $FM$, $FM'$, $SDW$, $EI$, $EI'$, and $CDW$.
The idea of the mean field approximation\cite{WenQFT} is to guess a trial ground state which can minimize the total energy of the many-body system. With a given trial ground state, the original Hamiltonian can be approximated by an effective quadratic mean field Hamiltonian which can be solved by self-consistent diagonalizing. 

The procedure will be briefly shown in the following. First, let the mean field Hamiltonian be   
\begin{equation*}
H_{MF}= H_0 +H_{pair}
\end{equation*}
$H_0$ is the free Hamiltonian given in Eq. \eqref{H0}, and
\begin{equation*}
H_{pair}=\sum_{\alpha\alpha';kk'}\Delta_{\alpha\alpha';\sigma\sigma'}(k,k')c^\dagger_{\alpha\sigma,k}f_{\alpha'\sigma',k'} + h.c.
\end{equation*}
For simplicity, since the coupling constants are independent of $k$ and $k'$, we assume $\Delta_{\sigma\sigma'}(k,k^\prime)\simeq\Delta_{\alpha\alpha';\sigma\sigma'}^{(0)}\delta^2(k-k^\prime)$ . Therefore, the pairing gap function $\Delta_{\alpha\alpha';\sigma\sigma'}^{(0)}$ (order parameter) is given by
\begin{equation*}\label{pairing}
\begin{split}
\frac{\Delta^{fm}_{\alpha\alpha';ss'}}{G_{fm}}&=\sum_{k} \langle c^\dagger_{\alpha s,k}\delta_{\alpha\alpha'}\otimes \sigma^z_{ss'} f_{\alpha's',k}\rangle\\
\frac{\Delta^{fm'}_{\alpha\alpha';ss'}}{G_{fm'}}&=\sum_{k} \langle c^\dagger_{\alpha s,k}\tau^z_{\alpha\alpha'}\otimes \sigma^z_{ss'} f_{\alpha's',k}\rangle\\
\frac{\Delta^{sdw}_{\alpha\alpha';ss'}}{G_{sdw}}&=\sum_{k} \langle c^\dagger_{\alpha s,k}\tau^x_{\alpha\alpha'}\otimes \sigma^z_{ss'} f_{\alpha's',k}\rangle\\
\frac{\Delta^{ei}_{\alpha\alpha';ss'}}{G_{ei}}&=\sum_{k} \langle c^\dagger_{\alpha s,k}\delta_{\alpha\alpha'}\otimes \delta_{ss'} f_{\alpha's',k}\rangle\\
\frac{\Delta^{ei'}_{\alpha\alpha';ss'}}{G_{ei'}}&=\sum_{k} \langle c^\dagger_{\alpha s,k}\tau^z_{\alpha\alpha'}\otimes \delta_{ss'} f_{\alpha's',k}\rangle\\
\frac{\Delta^{cdw}_{\alpha\alpha';ss'}}{G_{cdw}}&=\sum_{k} \langle c^\dagger_{\alpha s,k}\tau^x_{\alpha\alpha'}\otimes \delta_{ss'} f_{\alpha's',k}\rangle
\end{split}
\end{equation*}
where $G_j=\Gamma_j(0)$ is a linear combination of the bare coupling constants listed  in Table \ref{Gamma}. To find the trial ground state, a new quasi-particle (mixture of $c,v$ bands) is introduced which diagonalizes the mean field Hamiltonian,
\begin{eqnarray*}
\eta_{\alpha\sigma,k}&=v^{\alpha\alpha';\sigma\sigma'}_k c_{\alpha\sigma,k}-u^{\alpha\alpha';\sigma\sigma'}_k f_{\alpha'\sigma,k}\\
\lambda_{\alpha\sigma,k}&=u^{\alpha\alpha';\sigma\sigma'}_k c_{\alpha\sigma,k}+v^{\alpha\alpha';\sigma\sigma'}_k f_{\alpha'\sigma,k}
\end{eqnarray*}
Where $|u(k)|^2+|v(k)|^2=1$.
\begin{equation*}
\begin{split}
\{\eta_{\alpha\sigma}(k),\eta^\dagger_{\alpha'\sigma'}(k')\}&=\delta_{\alpha\alpha'}\delta_{\sigma\sigma'}\delta^2(k-k')\\
\{\lambda_{\alpha\sigma}(k),\lambda^\dagger_{\alpha'\sigma'}(k')\}&=\delta_{\alpha\alpha'}\delta_{\sigma\sigma'}\delta^2(k-k')
\end{split}
\end{equation*}
The other commutation relation are zero. Then, the diagonalized Hamiltonian can be written as
\begin{equation*}
H_{MF}=\sum_{\sigma,k}E(k)[\eta^\dagger_{\alpha\sigma}(k)\eta_{\alpha\sigma}(k)-\lambda^\dagger_{\alpha\sigma}(k)\lambda_{\alpha\sigma}(k)\;]
\end{equation*}
where, 
\begin{equation*}
E(k)=\sqrt{|\epsilon(k)|^2+|\Delta^{(0)}|^2},
\end{equation*}
and
\begin{equation*}
u^{\alpha\alpha';\sigma\sigma'}_k=\frac{1}{\sqrt2}(\delta_{\alpha\alpha'}\delta_{\sigma\sigma'})\left(1+\frac{\epsilon(k)}{E(k)}\right)^{\frac{1}{2}},
\end{equation*}
\begin{equation*}
v^{\alpha\alpha';\sigma\sigma'}_k=\frac{1}{\sqrt2}\left(\frac{\Delta_{\alpha\alpha';\sigma\sigma'}^{(0)}}{|\Delta^{(0)}_{\alpha\alpha';\sigma\sigma'}|}\right)\left(1-\frac{\epsilon(k)}{E(k)}\right)^{\frac{1}{2}},
\end{equation*}
$\epsilon(k)=|\epsilon_{\pm}(k)|$ is given by Eq. (\ref{dispersion})

Therefore, using a Hartree-Fock state as the trial ground state for the quasi-particles, we have $\prod_{\alpha,k}\lambda^\dagger_{\alpha\sigma}(k)|0\rangle=|\Psi_{\Delta^{(0)}}\rangle$, where $|0\rangle$ is the state with no particles (vacuum). To calculate $\Delta^{(0)}$, one can apply $|\Psi_{\Delta^{(0)}}\rangle$ to calculate the order parameter and obtain the gap equations,
\begin{equation*}
\Delta^{(0)}_{\alpha\alpha';\sigma\sigma'}=\sum_{k}\frac{G\Delta^{(0)}_{\alpha\alpha';\sigma\sigma'}}{\sqrt{|\epsilon(k)|^2+|\Delta^{(0)}_{\alpha\alpha';\sigma\sigma'}|^2}}
\end{equation*}
which are then solved self-consistently.  

\bibliography{QEBib}

\end{document}